\documentclass[useAMS,usenatbib]{mn2e}

\usepackage{graphicx}

\title{Optical properties of SMC X-ray binaries}

\author[M.J. Coe et al]{M.J.~Coe, W.R.T.~Edge, J.L.~Galache and V.A.~McBride  \\
       School of Physics and Astronomy, Southampton University, SO17 
1BJ, UK\\}
\begin{document}

\date{13 Aug 2004}

\pagerange{\pageref{firstpage}--\pageref{lastpage}} \pubyear{2004}

\maketitle

\label{firstpage}

\begin{abstract}

This work represent the first major study of the optical and IR
characteristics of the mass donor companions to the X-ray pulsars in
the Small Magellanic Cloud (SMC).  In this work several new
counterparts have been identified, and possible ones confirmed, as
companions to X-ray pulsars in the SMC giving a total of 34 such
objects now identified. In addition this work presents three new
binary periods and confirms two X-ray periods using optical data for
objects in this group. This homogeneous sample has been studied as a
group to determine important general characteristics that may offer
insight into the evolution of such systems. In particular, the
spectral class distribution shows a much greater agreement with those
of isolated Be stars, and appears to be in some disagreement with the
galactic population of Be stars in Be/X-ray binaries. Studies of the
long term optical modulation of the Be star companions reveal an
extremely variable group of objects, a fact which will almost
certainly make a major contribution to the pronounced X-ray
variability. The spatial distribution of these systems within the SMC
is investigated and strongly suggests a link between massive star
formation and the HI density distribution.  Finally, studies of the
circumstellar disk characteristics reveal a strong link with optical
variability offering important clues into the long-term stability of
such disks.

\end{abstract}


\section{Introduction and background}

The Be/X-ray systems represent the largest sub-class of massive X-ray
binaries.  A survey of the literature reveals that of the 115
identified massive X-ray binary pulsar systems (identified here means
exhibiting a coherent X-ray pulse period), most of the systems fall
within this Be counterpart class of binary.  The orbit of the Be star
and the compact object, presumably a neutron star, is generally wide
and eccentric.  X-ray outbursts are normally associated with the
passage of the neutron star close to the circumstellar disk (Okazaki
\& Negueruela, 2001). A recent review of these systems may be found in
Coe (2000).

A Be star is a non supergiant hot star, with a B spectral type, whose
spectrum has, or had at some time, one or more Balmer lines in
emission.  The optical light from a Be/X-ray binary is dominated by
the mass donor star in the blue end of the spectrum, but at the red
end there is normally a significant contribution from the
circumstellar disk. Long term optical observations provide valuable
insights into the behaviour of the circumstellar disk, and hence into
some of the details of the binary interactions within the system.

X-ray satellite observations have revealed that the SMC contains an
unexpectedly large number of High Mass X-ray Binaries (HMXB). At the
time of writing, 47 known or probable sources of this type have been
identified in the SMC and they continue to be discovered at a rate of
about 2-3 per year, although only a small fraction of these are active
at any one time because of their transient nature.  Unusually
(compared to the Milky Way and the LMC) all the X-ray binaries so far
discovered in the SMC are HMXBs, and equally strangely, only one
of the objects is a supergiant system, all the rest are Be/X-ray binaries.

The X-ray properties of many of the SMC Be/X-ray binaries were
presented in the works of Majid, Lamb \& Macomb (2004) and Haberl \&
Pietsch (2004). In those paper the authors summarised the known data
on the high energy emission from the accreting pulsars. Haberl \&
Pietsch (2004) also proposed several optical counterparts based upon
positional coincidences between the X-ray and optical catalogues. In
this work several different/new counterparts are identified and the
detailed optical properties of most of the sample of 48 objects are
presented. Comparisons with the galactic sample of such objects are
also made. Since the extinction and distance to the SMC is well
documented, this provides an excellent opportunity to study an
homogeneous population of objects in a different environment to that
of the galaxy.

A list of all known X-ray pulsars may be found in Table~\ref{tab2}. To
minimise confusion arising from long, similar source names only the
short SXP identities will be used in this paper. This identity is
created simply from the the acronym SXP ({\bf S}mall magellanic cloud
{\bf X}-ray {\bf P}ulsar) followed by the pulse period in seconds to
three significant figures.





\begin{table*}
\begin{center}
\begin{tabular}{|l|l|l|l|}
\hline
\textbf{Short ID}& 
\textbf{RA (2000)}& 
\textbf{Dec (2000)}& 
\textbf{Full or alternative name} \\
\hline
 & & & \\
\hline
SXP0.09& 
00:42:35.0& 
-73:40:30.0& 
AX J0043-737  \\

SXP0.72& 
01:17:05.2& 
-73:26:36.0& 
SMC X-1  \\

SXP0.92& 
00:45:35.0& 
-73:19:02.0& 
PSR J0045-7319 \\

SXP2.16& 
01:19& 
-73:12& 
XTE SMC2165 \\

SXP2.37& 
00:54:34.0& 
-73:40:43.0& 
SMC X-2  \\

SXP2.76& 
00:59:12.8& 
-71:38:44.0& 
RX J0059.2-7138  \\

SXP3.34& 
01:05:02.0& 
-72:11:00.0& 
AX J0105-722, RX J0105.1-7211, [MA93]1506 \\

SXP4.78& 
00:52& 
-72:19& 
XTE J0052-723 \\

SXP6.85& 
01:01& 
-72:43& 
XTE J0103-728 \\

SXP7.78& 
00:52:07.7& 
-72:25:43.7& 
SMC X-3  \\

SXP8.02& 
01:00:42.8& 
-72:11:32.0& 
CXOU J010042.8-721132  \\

SXP8.80& 
00:51:52.0& 
-72:31:51.7& 
RX J0051.8-7231,1E0050.1-7247, [MA93]506  \\

SXP9.13& 
00:49:13.6& 
-73:11:37.0& 
AX J0049-732, RX J00492-7311  \\

SXP15.3& 
00:52:15.3& 
-73:19:14.0& 
RX J0052.1-7319, [MA93]552 \\

SXP16.6& 
00:50& 
-73:16& 
XTE J0050-732 \\

SXP18.3&
00:55& 
-72:42& 
XTE SMC pulsar XTE J0055-727 \\

SXP22.1& 
01:17:40.5& 
-73:30:52.0& 
RX J0117.6-7330,X Nova92, [MA93]1845  \\

SXP31.0& 
01:11:09.0& 
-73:16:46.0& 
XTE J0111.2-7317 \\

SXP34.1& 
00:55:27.9& 
-72:10:58.0& 
CXOU J005527.9-721058, RXJ0055.4-7210 \\

SXP46.4& 
& 
& 
XTE SMC pulsar  \\

SXP46.6& 
00:53:53.8& 
-72:26:35.0& 
1WGA 0053.8-7226,XTE J0053-724 \\

SXP51.0& 
& 
& 
XTE SMC pulsar  \\

SXP59.0& 
00:54:57.4& 
-72:26:40.3& 
RX J0054.9-7226,XTE J0055-724, [MA93]810 \\

SXP74.7& 
00:49:04.6& 
-72:50:53.0& 
RX J0049.1-7250, AX J0049-729 \\

SXP82.4& 
00:52:09& 
-72:38:03& 
XTE J0052-725, MACS J0052-726\#004 \\

SXP89.0& 
& 
& 
XTE SMC pulsar  \\

SXP91.1& 
00:50:55.8& 
-72:13:38.0& 
AX J0051-722 ,RX J0051.3-7216, [MA93]413 \\

SXP95.2& 
00:52& 
-72:45& 
XTE SMC95 pulsar \\

SXP101& 
00:57& 
-73:25& 
AX J0057.4-7325, RX J0057.3-7325 \\

SXP138& 
00:53:23.8& 
-72:27:15.0& 
CXOU J005323.8-722715, [MA93]667 \\

SXP140& 
00:56:05.2& 
-72:22:00.0& 
XMMU J005605.2-722200,2E0054.4-7237, [MA93]904 \\

SXP144& 
& 
& 
XTE SMC pulsar  \\

SXP152& 
00:57:50.3& 
-72:07:56.0& 
CXOU J005750.3-720756, [MA93]1038 \\

SXP165& 
& 
& 
XTE SMC pulsar \\

SXP169& 
00:52:54.0& 
-71:58:08.0& 
XTE J0054-720, AX J0052.9-7158, [MA93]623 \\

SXP172& 
00:51:50.0& 
-73:10:40& 
AX J0051.6-7311, RX J0051.9-7311, [MA93]504 \\

SXP202& 
00:50:20.8& 
-72:23:16.0& 
1XMMU J005920.8-722316 \\

SXP264& 
00:47:23.7& 
-73:12:27.0& 
XMMUJ004723.7-731226, RXJ0047.3-7312,[MA93]172 \\

SXP280& 
00:58:08.0& 
-72:03:30.0& 
AX J0058-72.0, [MA93]1036  \\

SXP293& 
00:50& 
-73:06& 
XTE J0051-727  \\

SXP304& 
01:01:02.7& 
-72:06:58.0& 
CXOU J010102.7-720658, [MA93]1240,RXJ0101.0-7206 \\

SXP323& 
00:50:44.8& 
-73:16:06.0& 
AX J0051-73.3,RXJ0050.7-7316, [MA93]387 \\

SXP348& 
01:03:13.0& 
-72:09:18.0& 
SAX J0103.2-7209, RX J0103-722, [MA93]1367 \\

SXP452& 
01:01:20.5& 
-72:11:18.0& 
RX J0101.3-7211, [MA93]1257 \\

SXP504& 
00:54:55.6& 
-72:45:10.0& 
CXOU J005455.6-724510,RXJ0054.9-7245,AXJ0054.8-7244, [MA93]809 \\

SXP565& 
00:57:36.2& 
-72:19:34.0& 
CXOU J005736.2-721934, [MA93]1020 \\

SXP701& 
00:55:17.9& 
-72:38:53.0& 
XMMU J005517.9-723853 \\

SXP756& 
00:49:40.0& 
-73:23:17.0& 
AX J0049.4-7323, RX J0049.7-7323, [MA93]315 \\

\end{tabular}

\caption{List of known X-ray pulsars in the SMC. The [MA93] object
numbers come from Meyssonnier \& Azzopardi, 1993. Detailed lists of
published references
for many of the sources may be found in Haberl \& Pietsch (2004).}
\label{tab2}
\end{center}
\end{table*}


\section{Photometry and optical identification}

\subsection{SAAO data}

Optical photometric observations were taken from the SAAO 1.0m
telescope on the nights of 14 \& 15 December 2003. The data were
collected using the Tek8 CCD giving a field of $\sim$6x6 arc minutes
and a pixel scale of 0.6 arcsec/pixel.  Observations were made through
standard Johnson-Cousin filters plus an H$\alpha $ filter.

\subsection{Other photometric data}

Standard B, V and I photometric data were obtained from the OGLE II
survey of the SMC (Udalski et al., 1998).  Driftscan observations were
carried out on the 1.3m Warsaw telescope at the Las Campanas
Observatory, Chile.  Data of 11 fields covering the most dense parts
of the SMC were collected with the SITe 2048$\times$2048 CCD with a pixel
size of 24$\mu$m.

In cases where data from the OGLE II project were not available, B, V
and I magnitudes from the Magellanic Cloud Photometric Survey
(Zaritsky et al., 2002) have been presented in Tables 2 \& 3.  Driftscan
observations of the SMC were obtained using the Las Campanas Swope
telescope and a 2K CCD, with U, B, V and I magnitudes converted to the
standard Johnson-Kron-Cousins photometric system.





\begin{table*}
\begin{center}
\begin{tabular}{cccccccccccccc}
\hline
Short name& V& $\delta $V& B-V& $\delta $(B-V)& V-I& $\delta $(V-I)&
J& $\delta $J& J-K& $\delta $(J-K)& V-K& $\delta $(V-K)& Ref \\
\hline
SXP0.09& 
\multicolumn{13}{c}{Optical counterpart not yet identified} \\
\hline
SXP0.72& 
13.15& 
0.10& 
-0.14& 
0.15& 
-0.02& 
0.16& 
13.45& 
0.02& 
-0.03& 
0.05& 
-0.32& 
0.11& 
Z \\
\hline
SXP0.92& 
16.18& 
0.02& 
-0.21& 
0.03& 
0.15& 
0.03& 
16.60& 
0.16& 
$<0.31$& 
& 
$<-0.11$& 
& 
O \\
\hline
SXP2.16& 
\multicolumn{13}{c}{Optical counterpart not yet identified} \\
\hline
SXP2.37& 
16.64& 
0.04& 
0.06& 
0.06& 
0.15& 
0.07& 
& 
& 
& 
& 
& 
& 
Z \\
\hline
SXP2.76& 
14.01& 
0.08& 
0.06& 
0.09& 
-0.02& 
0.09& 
14.04& 
0.04& 
$<0.38$& 
& 
$<0.36$& 
& 
Z \\
\hline
SXP3.34& 
15.63& 
0.03& 
-0.01& 
0.05& 
0.03& 
0.05& 
15.76& 
0.07& 
0.32& 
0.19& 
0.19& 
0.18& 
O \\
\hline
SXP4.78& 
\multicolumn{13}{c}{Optical counterpart uncertain - see Section 5}\\
\hline
SXP6.85& 
\multicolumn{13}{c}{Optical counterpart not yet identified} \\
\hline
SXP7.78& 
14.91& 
0.02& 
0.00& 
0.03& 
0.08& 
0.05& 
14.82& 
0.07& 
0.37& 
0.11& 
0.46& 
0.09& 
Z \\
\hline
SXP8.02& 
18.09& 
0.02& 
-0.32& 
0.03& 
~& 
~& 
~& 
~& 
& 
& 
& 
& 
V \\
``& 
18.01& 
0.02& 
-0.12& 
0.04& 
-0.12& 
0.05& 
~& 
~& 
& 
& 
& 
& 
O \\
\hline
SXP8.80& 

14.87& 
0.12& 
-0.27& 
0.13& 
-0.04& 
0.18& 
14.47& 
0.03& 
0.21& 
0.07& 
0.61& 
0.14& 
O \\

\hline
SXP9.13& 
16.51& 
0.02& 
0.10& 
0.04& 
0.30& 
0.03& 
16.11& 
0.09& 
$<0.22$& 
& 
$<0.63$& 
& 
O \\
\hline
SXP15.3& 
14.67& 
0.04& 
-0.01& 
0.05& 
0.15& 
0.06& 
14.40& 
0.03& 
0.29& 
0.07& 
0.56& 
0.08& 
O \\
\hline
SXP16.6& 
\multicolumn{13}{c}{Optical counterpart not yet identified} \\
\hline
SXP18.3& 
\multicolumn{13}{c}{Optical counterpart not yet identified} \\
\hline
SXP22.1& 
14.18& 
0.03& 
-0.04& 
0.04& 
0.09& 
0.05& 
13.98& 
0.03& 
0.27& 
0.06& 
0.46& 
0.06& 
Z \\
\hline
SXP31.0& 
15.52& 
0.03& 
-0.10& 
0.04& 
0.23& 
0.04& 
15.10& 
0.06& 
0.19& 
0.14& 
0.62& 
0.13& 
Z \\
\hline
SXP34.1&
16.78&
0.03&
-0.12&
0.04&
-0.13&
0.05&
&
&
&
&
&
&
Z \\
\hline
SXP46.4& 
\multicolumn{13}{c}{Optical counterpart not yet identified} \\
\hline
SXP46.6& 
14.72&
0.03&
-0.07&
0.03&
0.14&
0.03&
14.41&
0.04&
0.42&
0.08&
0.72&
0.07&
Z \\
\hline
SXP51.0& \multicolumn{13}{c}{Optical counterpart not yet identified} \\
\hline
SXP59.0& 
15.28& 
0.01& 
-0.04& 
0.02& 
0.15& 
0.03& 
15.18& 
0.05& 
0.17& 
0.14& 
0.27& 
0.13& 
O \\
``& 
15.64& 
0.02& 
-0.21& 
0.03& 
0.20& 
0.03& 
~& 
~& 
& 
& 
& 
& 
V \\
\hline
SXP74.7& 
16.92& 
0.06& 
0.09& 
0.10& 
0.24& 
0.07& 
16.35& 
0.12& 
0.76& 
0.26& 
1.33& 
0.24& 
O \\
\hline
SXP82.4& 
15.02&
0.02&
0.14&
0.03&
0.3&
0.03&
14.63&
0.03&
0.42&
0.07&
0.81&
0.07&
O \\
\hline
SXP89.0&
\multicolumn{13}{c}{Optical counterpart not yet identified} \\
\hline
SXP91.1& 
15.06& 
0.06& 
-0.08& 
0.06& 
0.20& 
0.07& 
14.84& 
0.04& 
0.47& 
0.08& 
0.69& 
0.09& 
Z \\
\hline
SXP95.2& \multicolumn{13}{c}{Optical counterpart not yet identified} \\
\hline
SXP101& \multicolumn{13}{c}{Optical counterpart not yet identified} \\
\hline
SXP138& 
16.19& 
0.12& 
-0.09& 
0.12& 
0.05& 
0.26& 
16.27& 
0.11& 
$<-0.23$& 
& 
$<-0.31$& 
& 
Z \\
\hline
SXP140& 
15.88& 
0.03& 
-0.04& 
0.03& 
-0.11& 
0.05& 
15.85& 
0.09& 
$<1.75$& 
& 
$<1.78$& 
& 
Z \\
``& 
16.51& 
0.02& 
-0.25& 
0.03& 
0.65& 
0.03& 
~& 
~& 
& 
~& 
& 
& 
V \\
\hline
SXP144& \multicolumn{13}{c}{Optical counterpart not yet identified} \\
\hline
SXP152& 
15.69& 
0.03& 
-0.03& 
0.12& 
0.22& 
0.04& 
15.44& 
0.05& 
0.65& 
0.13& 
0.90& 
0.12& 
Z \\
\hline
SXP165& \multicolumn{13}{c}{Optical counterpart not yet identified} \\
\hline
SXP169& 
15.53& 
0.02& 
-0.05& 
0.04& 
0.19& 
0.12& 
15.38& 
0.05& 
0.42& 
0.14& 
0.57& 
0.13& 
Z \\
\hline
SXP172& 
14.45& 
0.02& 
-0.07& 
0.02& 
0.09& 
0.02& 
14.43& 
0.03& 
0.26& 
0.08& 
0.28& 
0.07& 
O \\
\hline
\end{tabular}

\caption{Table of photometric values - Part I. In the final column
the symbol indicates the source of the optical photometry: an O symbol
refers to OGLE data (Udalski et al., 1998), the Z symbol
refers to Zaritsky et al (2002), and the V symbol refers to data
presented for the first time in this work from observations carried
out using the SAAO 1.0m telescope. All $\delta$ values indicate the
errors or limits on the value of that parameter. The IR data come from 2MASS.}

\label{tab:photom1}
\end{center}
\end{table*}

\begin{table*}
\begin{center}

\begin{tabular}{cccccccccccccc}
\hline
Short name& 
V& 
$\delta $V& 
B-V& 
$\delta $(B-V)& 
V-I& 
$\delta $(V-I)& 
J& 
$\delta $J& 
J-K& 
$\delta $(J-K)& 
V-K& 
$\delta $(V-K)& 
Ref \\
\hline
SXP202& 
15.06& 
0.02& 
-0.24& 
0.03& 
0.23& 
0.03& 
~& 
~& 
& 
& 
& 
& 
V \\
``& 
14.83& 
0.02& 
-0.07& 
0.01& 
0.14& 
0.02& 
14.73& 
0.03& 
0.17& 
0.10& 
0.27& 
0.09& 
O \\
\hline
SXP264& 
15.85& 
0.01& 
0.00& 
0.01& 
0.23& 
0.01& 
15.53& 
0.12& 
$<0.91$& 
& 
$<1.24$& 
& 
O \\
\hline
SXP280& 
15.65& 
0.03& 
-0.12& 
0.04& 
0.13& 
0.04& 
15.44& 
0.06& 
0.56& 
0.12& 
0.77& 
0.11& 
Z \\
\hline
SXP293& 
\multicolumn{13}{c}{Optical counterpart not yet identified} \\
\hline
SXP304& 
15.72& 
0.01& 
-0.04& 
0.02& 
0.16& 
0.02& 
15.54& 
0.06& 
0.00& 
0.19& 
0.18& 
0.18& 
O \\
\hline
SXP323& 
15.44& 
0.04& 
-0.04& 
0.05& 
0.17& 
0.06& 
15.30& 
0.05& 
0.49& 
0.12& 
0.64& 
0.12& 
O \\
\hline
SXP348& 
14.79& 
0.01& 
-0.09& 
0.01& 
0.13& 
0.04& 
~& 
~& 
& 
& 
& 
& 
O \\
\hline
SXP452& 
15.49& 
0.02& 
-0.07& 
0.05& 
0.08& 
0.03& 
15.39& 
0.06& 
0.31& 
0.14& 
0.40& 
0.13& 
O \\
\hline
SXP504& 
14.99& 
0.01& 
-0.02& 
0.01& 
0.20& 
0.01& 
14.77& 
0.04& 
0.37& 
0.08& 
0.58& 
0.07& 
O \\
\hline
SXP565& 
15.97& 
0.02& 
-0.02& 
0.04& 
0.26& 
0.03& 
15.74& 
0.07& 
0.41& 
0.21& 
0.64& 
0.20& 
O \\
\hline
SXP701& 
16.01& 
0.03& 
0.08& 
0.05& 
0.30& 
0.04& 
15.42& 
0.05& 
0.54& 
0.13& 
1.13& 
0.13& 
O \\
\hline
SXP756& 
14.98& 
0.02& 
0.05& 
0.03& 
0.25& 
0.04& 
14.55& 
0.04& 
0.25& 
0.08& 
0.67& 
0.07& 
O \\
``& 
15.22& 
0.02& 
-0.06& 
0.03& 
0.57& 
0.03& 
~& 
~& 
& 
& 
& 
& 
V \\
\hline
\end{tabular}

\caption{Table of photometric values - Part II. In the final column
the symbol indicates the source of the optical photometry: an O symbol
refers to OGLE data Udalski et al 1998), the Z symbol
refers to Zaritsky et al (2002), and the V symbol refers to data
presented for the first time in this work from observations carried
out using the SAAO 1.0m telescope. All $\delta$ values indicate the
errors or limits on the value of that parameter. The IR data come from 2MASS.
}

\label{tab:photom2}
\end{center}
\end{table*}


\subsection{IR magnitudes}

The infrared magnitudes are from the 2MASS All-Sky Catalog of Point
Sources. 2MASS uses two automated 1.3 m telescopes, one at
Mt. Hopkins, USA, and one at CTIO, Chile. The sky can be observed
simultaneously in three near-infrared bands: J, H and K using a
three-channel camera on each telescope with each channel consisting of
a 256x256 array of HgCdTe detectors.

\subsection{Summary of data}

In order to indentify previously unknown optical counterparts to the
SMX X-ray pulsars a star had to satisfy as many of the following 
conditions as possible:

\begin{enumerate}

\item It should lie within, or very close to the X-ray error circle.

\item It should exhibit H$\alpha$ in emission.

\item Its optical colours should match those expected for a O or B
type star in the SMC.

\item If possible it should also exhibit significant optical
variability in the OGLE or MACHO data.

\item If possible it should be in the 2MASS catalogue.

\end{enumerate}

A summary of all the photometric magnitudes and colours of the SXP
objects is presented in Table~\ref{tab:photom1} and
Table~\ref{tab:photom2}. The values have been collected from the
published literature (see references cited in tables) and from
observations carried out during this work at SAAO. In some cases the
photometric magnitudes for a particular object may differ by 0.2 - 0.3
magnitudes. This is perhaps not too surprising given the known
variability of Be stars - see Section 4 below for a further evidence
for such changes.

\begin{figure}\begin{center}
\includegraphics[width=85mm]{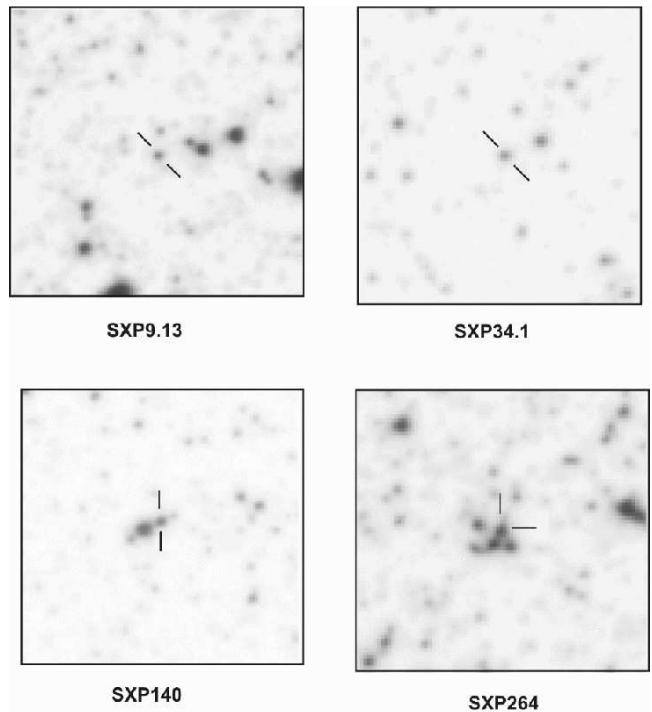}
\caption{Finding charts in the R band 
for selected objects. Each field is 1$\times$1
arcminute, north is up, east to the left.}
\label{fig:finders}
\end{center}
\end{figure}

Finding charts are shown in Figure~\ref{fig:finders} for two new
identifications where the optical counterpart is
not a previously catalogued object. In addition, the two other finding
charts (SXP140 \& SXP264) show objects
in crowded fields where the counterpart may not be immediately
obvious. In each case the charts show an R band image of
size 1$\times$1 arcminutes with the counterpart indicated.

\section{H$\alpha$ spectroscopy}

Spectroscopic observations in H$\alpha $ of possible optical
counterparts were made using the SAAO 1.9m telescope. A 1200 lines
mm$^{ - 1}$ reflection grating blazed at 6800{\AA} ~was used with the
SITe CCD which is effectively 266$\times$1798 pixels in size, creating a
wavelength coverage of 6160{\AA} to 6980{\AA}. The intrinsic
resolution in this mode was 0.42{\AA}/pixel.

\begin{table}
\begin{center}
\begin{tabular}{|l|c|l|l|l|}
\hline
\textbf{Object}& 
\textbf{H$\alpha$ profile}& 
\textbf{EW}& 
\textbf{Error on}& 
\textbf{Date of} \\

 & 
\textbf{}& 
\textbf{(\AA)}& 
\textbf{EW (\AA)}& 
\textbf{observation} \\
\hline
 & 
& 
& 
& 
 \\
  
SXP0.72& 
D& 
-2.5& 
0.2& 
031108 \\
  
SXP2.37& 
S& 
-7.4& 
0.4& 
990111 \\
  
SXP2.76& 
S& 
-20& 
0.5& 
031107 \\
  
SXP3.34& 
S& 
-54& 
1.4& 
021215 \\
  
  
SXP7.78& 
S& 
-12.4& 
0.6& 
031106 \\
    
SXP8.80& 
S& 
-10.2& 
0.6& 
031106 \\

SXP9.13& 
S& 
-29.6& 
1.5& 
031106 \\
  
SXP15.3& 
S& 
-18.4& 
0.5& 
031106 \\
  
SXP22.1& 
S& 
-19.6& 
0.6& 
031104 \\
  
SXP31.0& 
S& 
-22.8& 
0.7& 
031108 \\
  
SXP46.6& 
D& 
-21.9& 
0.7& 
031107 \\
  
SXP59.0& 
S& 
-12.5& 
0.7& 
031107 \\
  
SXP74.7& 
S& 
-11.1& 
1.1& 
031106 \\
  
SXP91.1& 
S& 
-21.9& 
0.7& 
031106 \\
  
SXP152& 
S& 
-19.8& 
1& 
031107 \\
  
SXP169& 
S& 
-24.5& 
0.8& 
031106 \\
  
SXP172& 
S& 
-13.1& 
0.5& 
011107 \\
  
SXP264& 
S& 
-31.6& 
0.9& 
031105 \\
  
SXP280& 
S& 
-37.3& 
0.9& 
031107 \\
  
SXP304& 
S& 
-55.2& 
1.3& 
031107 \\
  
SXP323& 
S& 
-24.5& 
0.7& 
031106 \\
  
SXP348& 
S/D& 
-17.5& 
0.7& 
031107 \\
  
SXP452& 
S& 
-11.1& 
0.8& 
011109 \\
  
SXP565& 
D& 
-28& 
2& 
031107 \\
  
SXP756& 
S& 
-22.9& 
0.7& 
011107 \\
  
\end{tabular}

\caption{H$\alpha$ emission line measurements. In the second column
the letter S refers to a single shaped profile and D to a double or
split profile. The date of the observation is in  the format YYMMDD.}
\label{tab:ha}

\end{center}
\end{table}

\begin{figure}
\includegraphics[width=75mm,angle=-0]{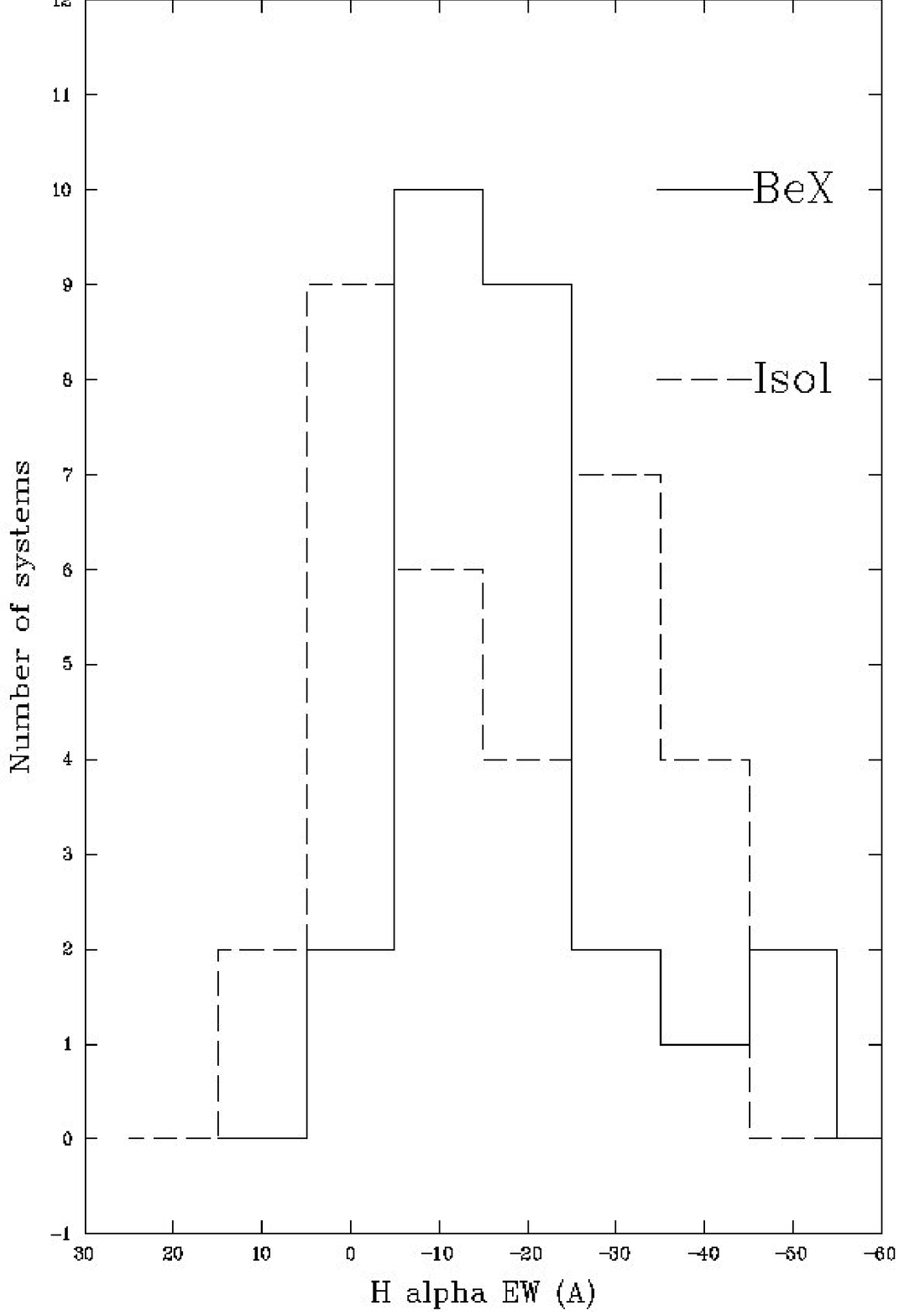}
\caption{Histogram of the H$\alpha$ equivalent widths for the systems
presented in this paper (solid line) and a sample of isolated Be stars
(dashed line - see text for references).}
\label{fig:ha}
\end{figure}

All our measurements of the strength of the H$\alpha$ emission lines from these
systems are presented in Table~\ref{tab:ha}. An approximate estimate
of the shape of the line is given in this table (single or double
profile), however the faintness of the objects for the telescope used
means that some small splittings of the line profiles 
may have gone undetected. A histogram of the H$\alpha$ equivalent
widths is presented in Figure~\ref{fig:ha} from which it can be seen
that the overwhelmingly most common EW values lie in the -10\AA ~to
-20\AA ~range. Also shown in this figure is a comparison sample of
galactic isolated Be stars. 
The data on isolated Be stars are taken
from the Be star surveys of Dachs \& Wamsteker (1982) and Ashok et al
(1984). See Section 6.4 for a discussion on this comparison.

Though H$\alpha$ profiles of Be stars are known to vary between single and
double peaks (see, for example, Reig et al., 1997), when in the double
state the separation of the peaks provides information on the mean
circumstellar disk velocities. 
A detailed analysis of the four H$\alpha$ spectra that are resolved in
these data revealed the following results on these objects.

SXP0.72 shows a clear double-peaked profile. Assuming a gaussian
profile for the peaks, we find the FWHM of each peak to be
$5.5\pm0.4$\AA, with a separation of 10.2\AA ~between the peaks. This
separation implies a radial velocity of the disk material of $\sim$230
km/s. The V/R ratio is $\sim$0.71.

Using the same tools we find that SXP46.6 shows a close double-peaked
profile with a FWHM of $5.8\pm0.3$\AA, a radial velocity of $\sim$100
km/s and V/R = $\sim$0.92.

SXP349 shows a highly disproportionate double profile which results in
a V/R = $\sim$1.76; the FWHM of the peaks is $4.1\pm0.2$\AA ~and the
radial velocity is also $\sim$100 km/s.

\begin{figure}\begin{center}
\includegraphics[width=85mm]{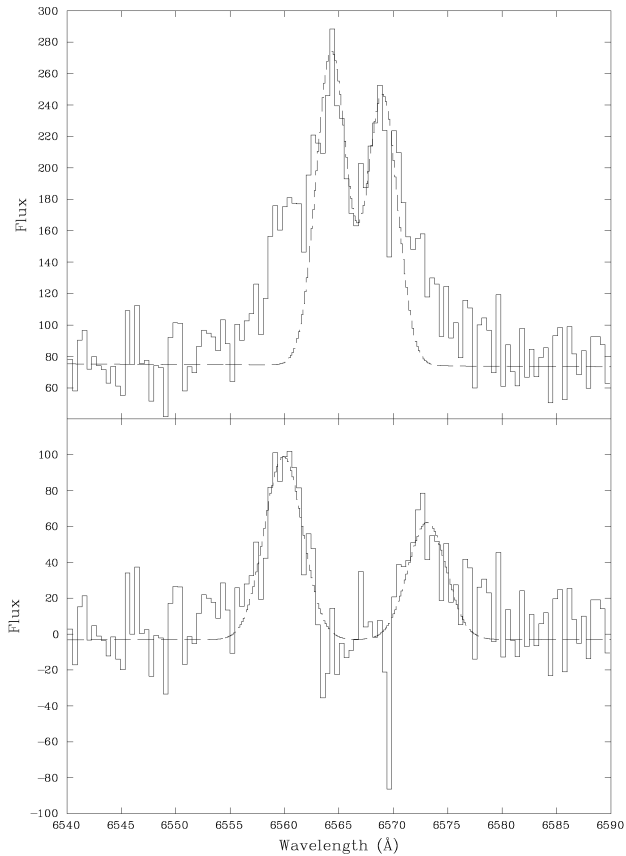}
\caption{The double H$\alpha$ profiles of SXP565. The upper figure
shows the raw data and the fit to the innermost pair of lines. The
lower panel shows the result of subtracting the above fit from the raw data.
}
\label{fig:spec}
\end{center}
\end{figure}

SXP564 has a more complicated profile than the previous sources,
showing 4 peaks (see Figure~\ref{fig:spec}). In the top panel a double
peak profile has been fitted with a FWHM of 3.3\AA ~and a separation of
4.7\AA ~between the peaks (providing a radial velocity of $\sim$105
km/s and V/R = $\sim$1.16). The resulting curve was then subtracted
from the observed data (see bottom panel), resulting in clear evidence
for two other peaks. A fit to this subtracted data results in a FWHM =
$4.1\pm0.4$\AA ~for the peaks, with a radial velocity of
$\sim$302 km/s and V/R = $\sim$1.56.  The two innermost peaks from
SXP564 would correspond to a high-density section in the outer edge of
the circumstellar disk, while the two outermost peaks could be a sign
that the star is undergoing a new period of mass ejection as the emmitting
region must be very close to the stellar surface.

\section{Periodicities and other variability}

The optical brightness of the companion stars to many X-ray
pulsars varies significantly over time and if these changes were
related to the orbit of the compact object, a study of the optical
lightcurves might yield information on the orbital period.
Conversely, the detection of an optical period could identify a
particular star as the optical counterpart from amongst a number
of otherwise similar candidates.

To this end, the MACHO lightcurves of all stars identified as
likely candidates were analyzed for periodicity. The selection of
possible counterparts differed with each pulsar: In some cases a
single star had been conclusively identified but in cases where
there was no obvious candidate the following selection criterea
were applied.

The first criterion, an optical (B-V) colour index, was obtained
by assuming that the stars being searched for were in the spectral
range B0V - B2V and that the extinction to the SMC was somewhere
in the range $0.08<E(B-V)<0.25$. The lower value comes from the
work of Schwering \& Israel (1991) and the upper value from direct
observations of similar Be systems (see, for example, Coe, Haigh
and Reig 2000) and includes a local contribution from the
circumstellar disk. The limits chosen were $-0.2<(B-V)<0.2$.

The second criterion was simply a cut-off in the V band magnitude
at 17.0, again set by assuming the same spectral class range as
above projected to the SMC through any reasonable amount of
interstellar and circumstellar absorption.

The third criterion, an infrared (J-K) colour index was determined
entirely from previous work. Since the circumstellar disk is the
single major contributor to the IR flux, the state of the disk
defines the size of the IR excess. Previously determined values
for (J-K) range from -0.1 up to 0.6, so a limit of $(J-K)<0.7$ was
chosen from these observations.

Schmidtke et al. (2004) report a similar study of the photometric
properties of SMC Be-neutron star systems but with the emphasis on
the detection of orbital periods in well established counterparts.

An attempt was also made to measure the gross variation in the
optical magnitudes by determining the standard deviation divided
by the mean($\sigma/\mu$). This was done using the generally
cleaner OGLE data. It is clear that some of these objects have a
longer variability than the length of the data run.

Those objects showing particularly striking variability are shown
in Figure~\ref{fig:lightcurves} which also includes the MACHO
lightcurves (covered in the following section). In order to avoid
cluttering the diagram, only the \textit{blue} data are shown. In
the case of SXP140, where no OGLE data exists, both the
\textit{blue} and \textit{red} curves are shown. The
transformation equations, given by Alcock et al. (1999), which
enable the instrumental magnitudes to be converted to approximate
$V$ and $R$ spectral values have been used.

\subsection{MACHO data}

In 1992 the  MAssive Compact Halo Objects project (MACHO) began a
survey of regular photometric measurements of several million
Magellanic Cloud and Galactic bulge stars (Alcock et al. 1993).

The MACHO data cover the period July 1992 to January 2000 and
consist of lightcurves in two colour bands described as
\textit{blue} and \textit{red}. \textit{Blue} is close to the
standard $V$ passband and \textit{Red }occupies a position in the
spectrum about halfway between $R$ and $I$ (Alcock et al. 1999).

These data were extensively analysed for periodicity using a
variety of period search algorithms. Initially all potential
optical counterparts were analyzed using the Nthalias binary
period detection program (T R Marsh, private communication). They were then
further examined using the Starlink PERIOD and the Grup d'Estudis
Astronòmics Analisis de Variabilidad Estelar (AVE) programs.
Finally all counterparts were examined with the Analysis of
Variance (AoV) periodogram devised by Aleksander
Schwarzenberg-Czerny (Schwarzenberg-Czerny. 1989). Apart from the
removal of non-physical observations (e.g those having magnitudes
of -99.99), the lightcurves have been used in their original form,
without being detrended or cleaned, so as to avoid the risk of
arbitrarily removing or introducing periodicity. The results are
shown in Table~\ref{tab1}. A number of likely orbital periods were
detected although in the majority of cases significant optical
periods were either not found at all or else only appeared in one
or two of the four data sets examined.

A period of around 29.6 days was found in a number of objects, and
for this reason it is likely to be an artefact. We have not
therefore included any mention of it in the following subsections.

\subsection{OGLE data}

The Optical Gravitational Lensing Experiment (OGLE) is a long term
project, started in 1992, with the main goal of searching for
dark matter with the microlensing phenomena.

Two sets of OGLE data, designated II and III, are available for
most of the objects discussed in this paper. Both show I-band
magnitudes using the standard system, however the more recent
OGLE-III data have not yet been fully calibrated to photometric
accuracy. OGLE II data points are earlier than MJD 52000, OGLE III
are later.

\begin{figure*}
\begin{center} \hbox{
\includegraphics[height=22cm]{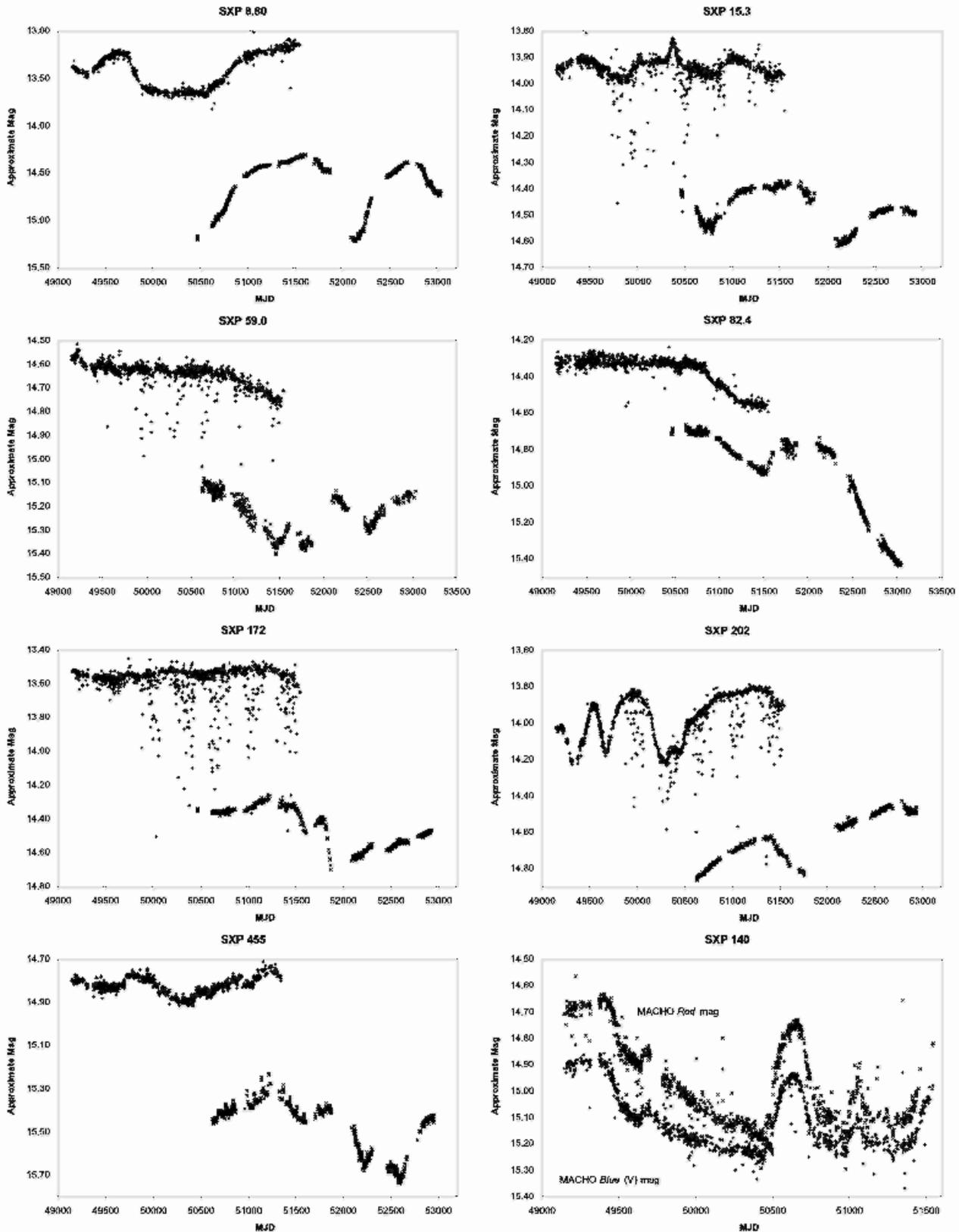}}
\end{center}
\caption{MACHO (top) and OGLE (bottom) Lightcurves. MACHO data
show approximate V-band magnitudes while OGLE data are in the
standard I-band. OGLE II data points are earlier than MJD 52000,
OGLE III are later and have not yet been fully calibrated to
precise photometric accuracy.In the case of SXP140, where only
MACHO data exists, \textit{Red} and \textit{Blue} data are shown.
MJD=Julian date -2450000.5.} \label{fig:lightcurves}
\end{figure*}




\begin{figure}
\begin{center}
\includegraphics[width=85mm]{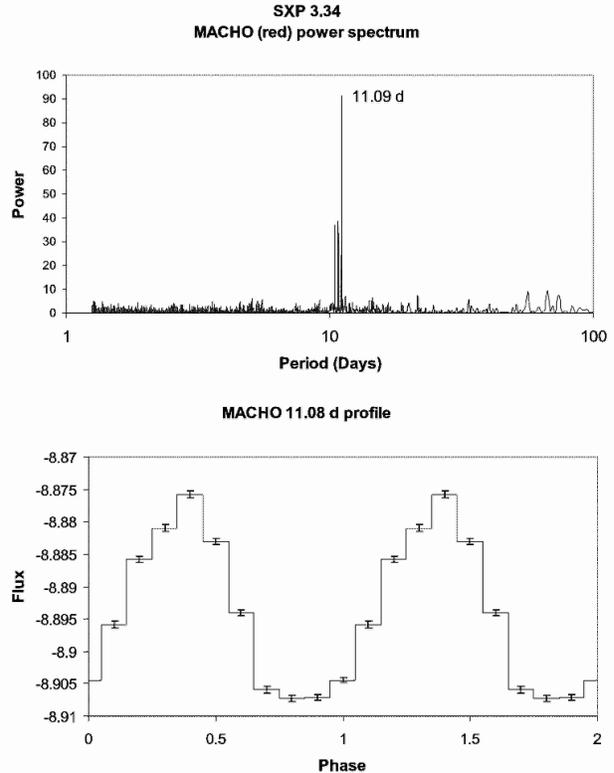}
\end{center}
\caption{SXP3.34 is one of relatively few objects to exhibit a
definite optical period. The power spectrum from the MACHO
\textit{red} data and pulse profile of the 11.08 day period are
shown. A triple structure is visible in the peak. Similar effects
are also visible at periods of $1/2$ and $1/3$.}
\label{fig:powerspectra}
\end{figure}

\begin{table*}
\begin{minipage}{150mm}
\begin{tabular}{|c|c|c|c|c|c|c|c|}
\hline Short ID& MACHO& OGLE II& OGLE III& OGLE& $P_{opt}$& Var.& Remarks \\
 & No.& No.& No.&SMC Field& (days)& $\sigma/\mu({\%})$& \\
%
%
\hline SXP0.92& & 11336& 19461& 4& & 0.9&
 \\
\hline SXP3.34& 206.16890.17& 102553& & 10& 11.09$\pm$0.01& 2&
 \\
\hline SXP4.78& 207.16146.9& & & & 23.9$\pm$0.06& &
 \\
\hline SXP7.78& 208.16088.4& & & & 44.6$\pm$0.2& &
 \\
\hline SXP8.02& & 42101& 35233& 9& & 5.4&
 \\
\hline SXP8.90& 208.16087.9& 85614& 4737& 6& 185$\pm$4& 22.2&
 \\
\hline SXP9.13& 212.15906.2446& 111490& 63223& 5& 91.5& 2.9&
 \\
\hline SXP15.3& 212.16075.13& 99923& 48026& 6& & 5.1&
 \\
\hline SXP46.6& 207.16202.7& & & & 152.4$\pm$2.4& &
red only \\
\hline SXP58.9& 207.16259.23& 70829& 35420& 7& & 7.8&
 \\
\hline SXP74.8& 208.15911.454& 65517& 55158& 5& & 2.4&
 \\
\hline SXP82.4& 208.16085.24& 77228& 12997& 6& & 19.7&
 \\
\hline SXP91.1& 208.16034.5& & & & 88.4$\pm$0.4& &
 \\
\hline SXP138& 207.16202.50& & & & & &
 \\
\hline SXP140& 207.16374.21& & & & & &
 \\
\hline SXP172& 212.16077.13& 22749& 44100& 6& & 9.6&
 \\
\hline SXP202& 207.16545.12& 151891& 4929& 8& & 4.7&
 \\
\hline SXP264& 212.15792.51& 116979& 45007& 4& & 2.4&
 \\
\hline SXP304& 206.16663.16& 47428& 12190& 9& & 0.7&
 \\
\hline SXP323& 212.16019.30& 180026& 48& 5& 0.708 and 1.695& 3.1&
 \\
\hline SXP349& 206.16776.17& 173121& & 9& & 2.7&
 \\
\hline SXP452& 205.16662.14& 89374& 34801& 9& 74.7& 7.3&
 \\
\hline SXP503& 207.16245.16& 47103& 36877& 7& & 1.1&
 \\
\hline SXP564& 207.16432.1575& 49531& 19293& 8& 95.3& 2.4&
 \\
\hline SXP701& 207.16313.35& 129062& 36015& 7& & 2&
 \\
\hline SXP755& 212.15960.12& 90506& 22903& 5& 394$\pm$2.3& 2.1&
 \\
\hline
%
%
\end{tabular}
\caption{Optical periods derived from MACHO, OGLE II and OGLE III
data. This table lists only those objects for which either MACHO
or OGLE data are available. Long-term variability
($\sigma/\mu({\%})$) was derived from OGLE data only.}
\label{tab1}
\end{minipage}
\end{table*}

\section{Discussion on individual sources}

\subsection{SXP3.34}

This X-ray pulsar is identified here with [MA93] 1506. This object
lies within the ASCA error circle (Yokogawa \& Koyama 1998) and
showns strong H$\alpha$ emission. In addition, this object was
found to have an unusually strong optical period of 11.09 days in
both types of data. This would be consistent with the Corbet
diagram (Corbet, 1986) if it was the orbital period. On the other
hand, the fact that it has an H$\alpha$ equivalent width of
-54\AA, is at variance with the relationship between orbital
period and EW established by Reig, Fabregat, \& Coe (1997) which
would imply a longer period. It is possible therefore that the true
binary period is considerably longer and that the one seen here
represents the the rotation period of part of the circumstellar
disk. The MACHO \textit{red} power spectrum and pulse profile are
shown in Figure~\ref{fig:powerspectra}. A curious triple structure
is visible in the peak and similar effects are seen in the OGLE
data and at periods of $1/2$ and $1/3$. There is no obvious
explanation for this. It is worth noting that apart from the enigmatic
periods in SXP323 (Coe et al., 2002), this may well be the shortest
known binary period for a Be/X-ray system.

\subsection{SXP4.78}

The possible counterparts to the XTE pulsar were discussed in Laycock
et al., (2003). Those authors suggested that the probable optical
counterpart was [MA93] 537, however the determination of the X-ray
position remains poor. It is therefore not possible to state with any
confidence exactly which object is the counterpart.  However, we note
that MACHO object 207.16146.9 at RA 0:52:38.39 Dec -72:21:2.7 was
found to have an optical period of $23.9\pm0.1$ days in both
colours. This object is identified with AzV129 which lies within the
XTE error box. An orbital period of $\sim$22 days would be anticipated
from the Corbet spin/orbit relationship.

\subsection{SXP7.78}

Corbet et al. (2004) give a probable binary period of $45.1\pm0.4$
days. An analysis of MACHO counterpart 208.16088.4 at RA: 00 52
05.5  Dec: -72 26 04.0 detected a period of $44.6\pm0.2$ days in
the \textit{red} lightcurve which contains large annual gaps in
the data.

\subsection{SXP8.80 (= SXP16.6?)}

Though previously identified with two different emission line
objects ([MA93] 506 by Haberl \& Pietsch (2004) and ``Star 1'' by
Israel et al., 1997) it is far from clear whether either of these
identifications are secure. In addition, there is a third star
(AzV111) lying about 15'' to the south of [MA93] 506 which also
shows H$\alpha$ emission and is inside both the ROSAT and Einstein
error circles. For the purpose of this paper we have used the star
identified by Haberl \& Pietsch since it lies in the centre of all
three independent X-ray positional determinations. A more precise
X-ray position is required to resolve this object. Corbet et al.,
(2004) have calculated a period of $28.0\pm0.3$ from RXTE X-ray
data. Laycock et al. (2004) have conjectured that this is the same
object as SXP16.6 for which they find an orbital period of
$189\pm18$ days based on X-ray data. Assuming object 208.16087.9
at RA: 00 51 53.0 Dec: -72 31 48.1 to be the MACHO counterpart, we
find a period of $185\pm4$ days in the \textit{red} data.
Significant periods were not found in the OGLE data.

\subsection{SXP9.13}

This source is identified with an H$\alpha$ bright object lying in the
revised ASCA error circle (Ueno et al., 2001). A finding chart is shown
in Figure~\ref{fig:finders}. In addition the position of this optical
object coincides with the ROSAT source RX J0049.2-7311. However, other
authors have proposed a second ROSAT source, RX J0049.5-7310, as the
correct identification for the X-ray pulsar (Filipovic et al (2000)
and Schmidtke et al (2004)).  Furthermore, Schmidtke et al (2004) find
an orbital period of 91.5 days for this second ROSAT source in the
MACHO data. However, this object lies well outside the revised X-ray
error circle presented in Ueno et al (2001) and consequently RX
J0049.2-7311 is presented here as a more likely counterpart to the
ASCA pulsar.

\subsection{SXP15.3}

The large peak in the MACHO data visible in
Figure~\ref{fig:lightcurves} coincides with a ROSAT HRI X-ray
detection on 19 Oct 1996 (MJD 50375)(Kahabka, 1999).

\subsection{SXP34.1}

This X-ray object was discovered from Chandra observations by Edge
et al. (2004). Its position was determined to $\sim$1 arcsec. The
nearest optical counterpart lies $\sim$2 arcsec away from the
X-ray position and is shown in Figure~\ref{fig:finders}. Its
optical colours are consistent with a B type star.

\subsection{SXP46.6}

Laycock et al. (2004) derive a period of $139\pm6$ days from the
X-ray data. Although we consider that the following is unlikely to
be the correct counterpart, we place on record that MACHO object
207.16202.30 at RA: 00 53 55.13 Dec: -72 26 31.8, which is 28
arcsec from the presumed star, was found to have a period of
$152.4\pm2.4$ days with a secondary period of $98\pm1$ days. This
object has approximate V and R magnitudes of 16.59 and 15.15
respectively, derived from the transformation equations given by
Alcock et al.(1999). The optical variability is clearly visible in
the lightcurves.

\subsection{SXP59.0}

Laycock et al. (2004) find a possible orbital period of $123\pm1$
days. We did not find any significant period in the optical data.

\subsection{SXP74.7}

Laycock et al. (2004) indicate a candidate orbital period of
$642\pm59$ days. No such period was found in any of the MACHO
candidates.

\subsection{SXP91.1}

The optical counterpart to this object was identified by Stevens,
Coe and Buckley (1999) with an $H\alpha$ emitting object
designated [MA 93]413 (Schmidtke et al., 2004). Laycock et al. (2004)
report a period of 115 days. Schmidtke et al. (2004) find an
optical period of 88.25 days using MACHO data, a finding confirmed
by our analysis of Object 208.16034.5 at RA: 00 50 57.0  Dec: -72
13 33.5. There are no OGLE data for this object.

We also place on record that MACHO object 208.16034.71,  RA: 00 51
00.75  Dec: -72 13 08.5, which is at a distance 40 arcsec from the
ROSAT position, was found to have an optical period of 115 days.
Even though the period seems to agree with the proposed X-ray period, 
the nominal ROSAT error of 10 arcsec would seem to eliminate this
object as a candidate.

\subsection{SXP140}

This is a difficult field - see Figure~\ref{fig:finders} - with
several sources blended together. Zaritsky et al. (2002) identify 3
objects, as does 2MASS, but there is probably at least one more object
in the blend. The object indicated in Figure~\ref{fig:finders} is
[MA93]904 - identified by Haberl \& Pietsch (2004) as the correct
counterpart based upon an XMM position. The SAAO data were carefully
deblended and the resulting magnitudes are presented in
Table~\ref{tab:photom1} for [MA93]904 
along with the Zaritsky et al. numbers. The
discrepancies between the two sets of data are noted but, until much
higher quality images of the field are obtained, they cannot be resolved.

\subsection{SXP172}

A significant proportion of the observations in the presumed MACHO
counterpart 212.16077.13 \textit{blue} data fall below the mean
curve in what appears to be an annual cycle. This is assumed to be
an artefact of the system. Laycock et al. (2004) derive a period
of $67\pm5$ days. No such period was found in the optical data.

\subsection{SXP202}

The presumed MACHO counterpart 207.16545.12 RA: 00 59 20.9  Dec:
-72 23 17.4 shows very high optical variability which is similar
to other optical counterparts but no significant optical period
was found.

\subsection{SXP323}

All four data sets consistently exhibited a period of 0.708 $\pm$
0.001 days and a slightly weaker period of 1.695 $\pm$ 0.003 days.
The former period has been reported and discussed by Coe et al.,
(2002).

\subsection{SXP452}

Schmidtke et al. (2004) report the discovery of an orbital period
at 74.7 days. We confirm a similar orbital period in MACHO object
205.16662.14 at RA: 01 01 20.6  Dec: -72 11 18.3 and conclude that
this is the same object as the one they report.

\subsection{SXP564}

Schmidtke et al. (2004) report an orbital period of 95.3 days. We
find the same period in MACHO object 207.16432.1575 RA: 00 57
35.98  Dec: -72 19 34.4. and conclude that this is the same
object.

\subsection{SXP756}

This object is unusual in that there are clearly visible peaks in
all the optical data, which coincide with X-ray detections,
confirming an orbital period of 394 days (Cowley \& Schmidtke,
2003, Coe \& Edge, 2004). The data points forming the short peaks
were excluded from the long-term variability calculation.

\section{Discussion of general properties}

\subsection{Long term optical variability}

Extensive analysis of up to four separate data sets for each
object, using a variety of period search algorithms, has not
generally been successful in determining the the orbital periods
of these X-ray binary stars. This finding is consistent with
previous observations of Galactic Be/X-ray binaries where no
modulation of the optical lightcurve at the known binary period
was detected, e.g.: A0535+26 (Clark et al. 1998 \& 1999). Although
a variety of periods could be detected, in most cases they were
ambiguous and did not occur in all the data sets. In some cases it
is possible that the wrong star has been identified as the optical
counterpart but there is good reason to believe that the majority
are correct. It is also possible, but unlikely, that the orbital
period may be marked by short disturbances which are not apparent
in the lightcurves and which are not detectable by period search
algorithms.

On the other hand, as can bee seen in
Figure~\ref{fig:lightcurves}, many of these systems undergo
significant long-term aperiodic fluctuations in luminosity. Roche
et al. (1997) have shown that, in the case of X Persei, a long
term variation in the V band photometry over 10 years represents a
phase change in the companion star from Be to OB and back. The
authors further demonstrate that the only source of variability in
the system is the circumstellar envelope and that during the OB
phase the star displays almost no variation. Negueruela et al.
(2001) use a similar interpretation to postulate a $\sim$3 year
cycle of disk loss and reformation in 4U0115+63/V635 Casiopeiae.

The objects studied in this paper, however, do not generally show
strong evidence of cyclic behavior of the sort seen in X Persei,
although SXP8.80 has a rather similar profile. SXP140, for
example, declines unevenly in brightness, by 0.53 mag over a
duration of $\sim1050$ days, then increases by 0.43 mag in 230
days and declines again in 220 days. If there is an underlying
cycle, it may be longer that the data set used in this study.

The most significant exception to the lack of evidence for orbital
period in the lightcurves is SXP755, which has several unusual
features. In the first place the orbital period had already been
established from the pattern of X-ray observations. Secondly, the
X-ray outbursts coincide with abrupt, spiky peaks which are
clearly visible in the optical data. These peaks are
non-sinusoidal and are not detectable by some types of periodic
analysis. The fact that this object has a long orbital period as
well as low variability suggests that the disk is in a relatively
stable configuration which is less prone to tidal disturbance by
the neutron star except at periastron passage.

\begin{figure}
\begin{center}
\includegraphics[width=60mm, angle=-90]{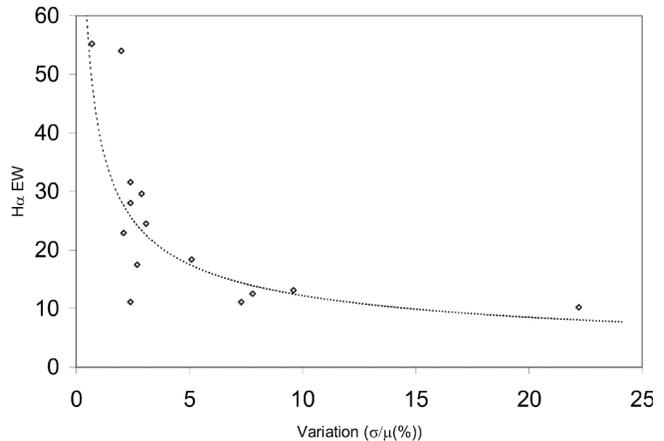}
\end{center}
\caption{Variability of the OGLE lightcurves is plotted against
H$\alpha$ equivalent width. The fitted curve shows an inverse
powerlaw relationship.} \label{fig:var}
\end{figure}

The possible relationship between the size and stability of the
circumstellar disk has been explored by comparing the disk size
with the star's variability. Figure~\ref{fig:var} plots the
equivalent width data from Table 4 against the optical variability
from Table 5. The fitted curve shows an inverse relationship of
the form: $EW \propto (Variability)^{-0.5}$.

Clearly the notion that low EW might be an indicator of high
variability is inconsistent with the principle that the photometric
state is correlated with disk size. Moreover the equivalent widths
in Table 4 are given for only a single point in time and it has
previously been stated that many Be stars are known to oscillate
between a Be and a B phase (Negueruela et al., 2001). It has also
been demonstrated that H$\alpha$ emissions can vary over very
short timescales (Clark et al. 1998). To test the validity of the
apparent relationship shown in Figure~\ref{fig:var}, it would
therefore be necessary to take spectroscopic measurements over a
period of time, in parallel with photometric ones. Long term
$H\alpha$ spectroscopic data is not available for the objects
studied here, if, however, a low photometric state were to be
interpreted as a disk-loss event then the low equivalent widths
seen in Table 4 might be expected to coincide with low points on
the lightcurves. We do not find such a correlation although the
curves generally show reddening in the brighter phases suggesting
a \textit{relative} enlargement of the disk. For example, where
the OGLE II (I) and macho \textit{blue} (V) overlap in SXP8.80,
the I band changes by 0.85 mag as compared with 0.5 for the V
band.

If the Figure~\ref{fig:var} relationship is genuine then the fact
that the variability appears to be greatest where the disk size is
least might be taken to suggest that it is intrinsic to the
companion star and not the disk. Although this seems partly to be
born out by the fact that it also appears strongly in the
\textit{blue} (V) part of the spectrum, such a conclusion is at
variance with strong evidence that optical variability in Be stars
results from changes in the disk, moreover Negueruela et al. have
shown that even when the disk is almost totally absent, the
presence of some circumstellar material contributes significantly
to the total brightness and hence variability.

\begin{figure}
\begin{center}
\includegraphics[width=60mm, angle=-90]{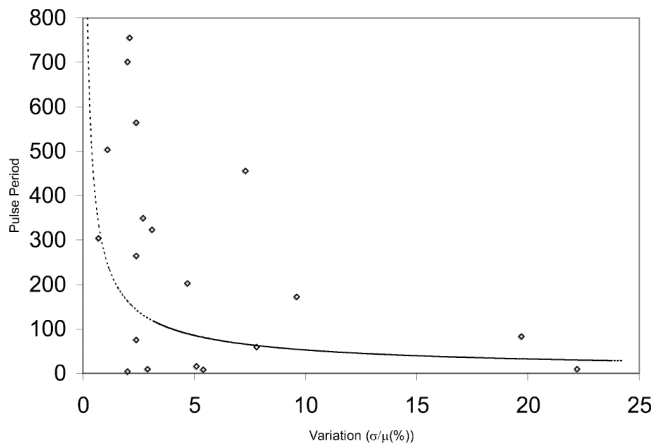}
\end{center}
\caption{Variability of the OGLE lightcurves is plotted against
pulse period. The fitted curve shows an inverse powerlaw
relationship.} \label{fig:var_pp}
\end{figure}

Reig, Fabregat, \& Coe (1997) have shown that there is also a
relationship between H$\alpha$EW and orbital period which is
attributed to increased truncation of the circumstellar disk with
shortened period. It might therefore be concluded from the
evidence of Figure~\ref{fig:var} that long term variability can be
related to short orbital period. Unfortunately there are not
enough orbital periods in this sample which are known with
sufficient confidence to test such a relationship directly, nor is
it immediately obvious what physical process would serve to
connect a binary period of a few tens of days with largely
aperiodic variations spanning years. A plot of pulse period
against variability (Figure~\ref{fig:var_pp}), however, appears to
show an inverse relationship of the kind seen in
Figure~\ref{fig:var}. This would be expected if both the Corbet
spin/orbit and the indicated variability/orbital period
relationships were valid and is indirect confirmation of the
latter.

\subsection{Spectral Class of counterparts}

In order to determine the approximate spectral class of the
counterparts listed in this work, the colour (B-V) of each object was
used.  To do this, the (B-V) colour was first corrected for the
extinction to the SMC of E(B-V) = 0.08 (Schwering \& Israel, 1991) and
then the resulting (B-V) used to identify the spectral class using the
data of Johnson (1966).  However, the resulting spectral class
distribution peaked a long way from the distribution of such objects
in our galaxy (see, for example, Negueruela 1998).  In order to adjust
the (B-V) colour scale for possible extra reddening (arising primarily
from the presence of the circumstellar disk), it was decided to
use the class determinations derived from independent blue spectra to
normalise the data presented here.  Fortunately, three of the objects
studied have had previously determined spectral classes based upon
such accurate blue band spectroscopy - see
Table~\ref{tab:cross}. Using these three objects, an average extra
(B-V) shift was calculated of (B-V)=--0.13 to bring them to their
correct spectral class. This shift was then applied globally to all
the objects.

\begin{table}
\begin{center}

\begin{tabular}{ccccc}
\hline
Source&Spectral&Predicted&Observed&Reference \\
&Class& (B-V) & (B-V)& \\
\hline
SXP31.0 &B0.5-B1 V & -0.19 & -0.10  & 1 \\
SXP323  &B0 III-V  & -0.22 & -0.04 & 2\\
SXP22.1 &B1-B2 III-V& -0.17 & -0.04 & 3\\
\hline
\end{tabular}

\caption{Spectral class previously determined from blue 
spectroscopy measurements (references are 1 - Covino et al., 2001, 2 -
Coe et al., 2002, 3 - Coe et al., 1998).  In each case the predicted
(B-V) is derived assuming the spectral class in column 2 and 
that E(B-V)=0.08 is the extinction to the
SMC (Schwering \& Israel, 1991). }

\label{tab:cross}
\end{center}
\end{table}

In each of the three reference objects listed in Table~\ref{tab:cross}
extra reddening had been reported as an individual effect - attributed
to either the circumstellar disk, or maybe due to column enhancements
at the objects locality. However, it is now very likely that this is a
general characteristic of all SMC Be/X-ray systems.

\begin{figure}\begin{center}
\includegraphics[width=65mm, angle =-90]{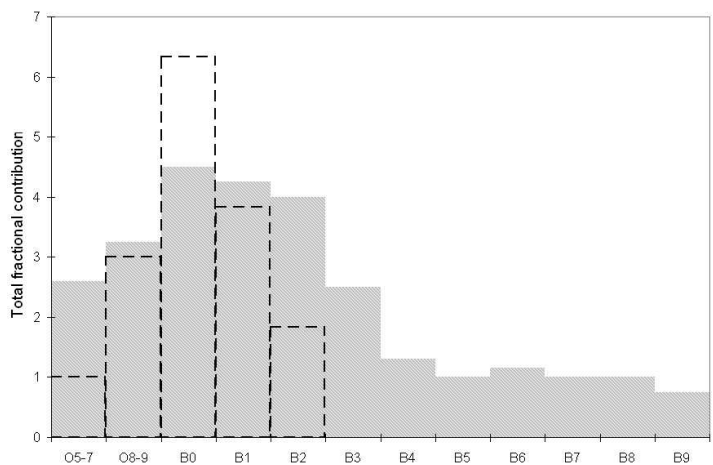}
\caption{Histogram of the spectral class distribution of the SMC
Be/X-ray binary counterparts (shaded) and those of galactic systems
(dashed line). See text for exact details on how this
figure was produced. }
\label{fig:sc}
\end{center}
\end{figure}

To allow for the uncertainty in the spectral class derived by this
method, each object was fractionally divided amongst the range of
classes consistent with its (B-V).  For example, the source SXP59.0 has a
B-V value of $-0.04\pm0.02$.  According to Johnson (1966) the
counterpart has a spectral type in the range from B0 to B2.
Consequently SXP59.0 has had one-third of its contribution allocated
to spectral type B0, another third to B1 and the final third to B2.
The resulting distribution is then the sum of all these fractional
contributions and is shown in Figure~\ref{fig:sc} as the shaded histogram.

For comparison, a histogram of galactic systems was also produced
(dashed line in Figure~\ref{fig:sc}). This distribution was determined
from the catalogue of Liu, van Paradijs \& van den Heuvel (2000) in a
similar manner to the SMC distribution. Namely, the objects selected
from the catalogue all have X-ray pulse periods and O/B type companions,
luminosity classes III or V. Again, where there was some
uncertainty in the spectral class the contribution of that source was
fractionally allocated. A total of 16 objects contribute to this
distribution.

This SMC distribution indicates that ``missing'' B3-B9 spectral class
counterparts to galactic Be/X-ray systems (Negueruela, 1998) may not
be similarily absent in the SMC systems. The distribution in the SMC
may well be intrinsically different to that in our own galaxy
reflecting a somewhat different evolutionary development. 
The much lower metallicity and the tidal interactions between the
clouds are two major factors that may well have played an important
role in promoting such a difference - though the models of van Bever \&
Vanbeveren 1997 question the importance of the metallicity factor.

\subsection{Locations of pulsars in the SMC}

\begin{figure}\begin{center}
\includegraphics[width=85mm]{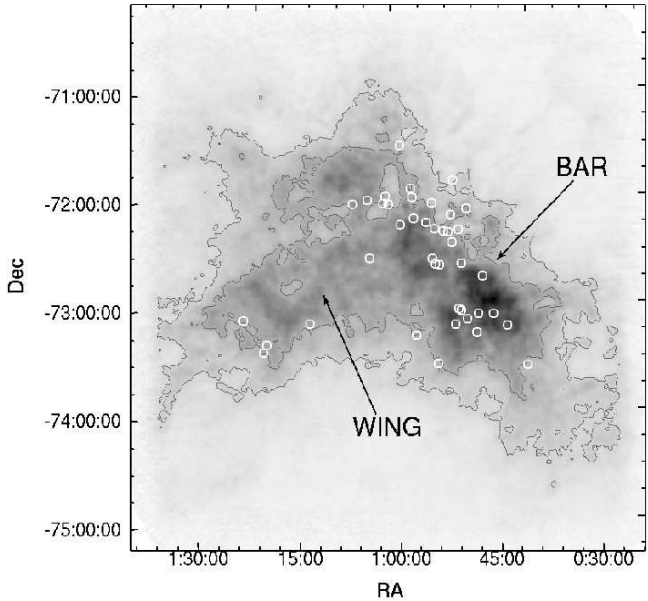}
\caption{Location of known X-ray pulsars superimposed on an HI map of
the SMC (Stanimirovic et al., 1999).
Grey-scale intensity range is 0 to
143$\times10^{20}$ atom cm$^{-2}$. Contour lines are at 19.9$\times10^{20}$
and 39.4$\times10^{20}$ atom cm$^{-2}$.
}
\label{fig:wingbar}
\end{center}
\end{figure}

\begin{figure}\begin{center}
\includegraphics[width=65mm,angle=90]{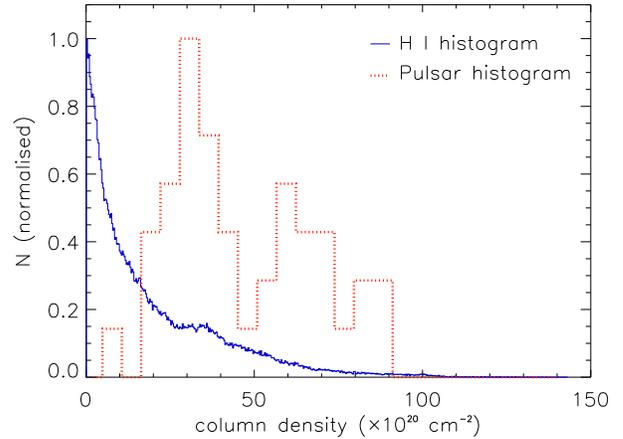}
\caption{Histogram of SMC HI intensity distributions 
(continuous line) and the corresponding histogram of HI columns at the
location of the X-ray pulsars (step curve).}
\label{fig:hi}
\end{center}
\end{figure}

The HI distribution of the SMC (from Stanimirovic et al., 1999) is
shown as a greyscale contour map in Figure~\ref{fig:wingbar};
superimposed on it are the 41 pulsars with well established
positions. It is therefore possible to extract the column density of
HI towards each of these pulsars and plot the resulting distribution 
as a histogram. This can
be seen in Figure~\ref{fig:hi}, together with the overall histogram of
HI density in the SMC; 25 bins were used for the pulsar data, 2000 for
the SMC HI.  In each case, the peak of the distribution was normalised
to 1.0. The figures indicate that the pulsars favour low/medium
HI densities. Over half the SMC population is found in the range of HI
density $\sim$(15-45) x ${10^{20}}$ ${cm^{-2}}$.  In this same region
of HI densities, the SMC histogram curve appears to deviate from an
exponential decay shape showing a clear dip. If these accurately
reflect the volume densities, there would be a strong suggestion that
high-mass star formation is particularly well suited to these
densities.

\begin{figure}\begin{center}
\includegraphics[width=85mm]{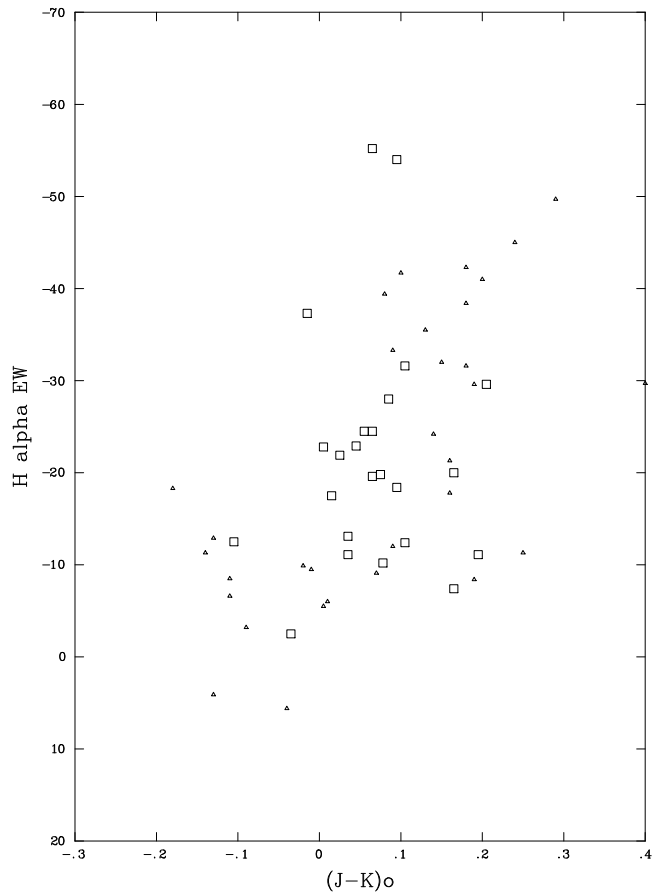}
\caption{Comparison of the properties of the SXP systems (square
symbols) with those of a sample of isolated galactic Be stars
(triangles) - see text for references.}
\label{fig:haew}
\end{center}
\end{figure}

It must be noted that the X-ray coverage by RXTE, Chandra and XMM of
the Wing of the SMC has been particularily sparse compared to that of
the Bar. Consequently the general distribution across different
components of the SMC is yet to be accurately determined. Since the
Wing and Bar are thought to be separated by up to 20kpc (Laney \&
Stobie, 1986) it wouldn't be surprising to find differing population
densities in the two components.

\subsection{Circumstellar disks in Be/X-ray binaries}

Figure~\ref{fig:haew} shows two measures of the size of the
circumstellar disk in these systems plotted against each other - the
H$\alpha$EW and the ``intrinsic'' IR colour $(J-K)_{o}$.  The data on
isolated Be stars are taken from the Be star surveys of Dachs \&
Wamsteker (1982) and Ashok et al (1984). The (J-K) colour for each SXP
object was corrected for all reddening effects by using the same
colour shift described in Section 6.2 to determine the spectral class.
This produced the ``intrinsic'' $(J-K)_{o}$ colours plotted in this
figure.  This figure reinforces the result presented earlier in
Figure~\ref{fig:ha} which suggests a similar range in disk sizes for the
isolated systems and the Be/X-ray binaries. Possibly 
the Be/X-ray systems do not extend down to the
lowest values, but this is probably a selection effect since we only
detect the more active systems. At the high value end the sample is too
sparse to draw any conclusions. In making plots like this one must
always be aware that the data are not taken contemporaneously, and
unusually high values in H$\alpha$ will not match the average IR
colours taken at some other time. Hence the Be/X-ray binaries with the
two highest H$\alpha$ EWs should almost certainly lie further to the
right on this diagram. 

As pointed out by Dachs et al. (1986) the disk inclination of these
systems is also likely to have an effect on the measured EWs. This
will have the result of spreading out the distribution in
Figure~\ref{fig:ha}. However, from our H$\alpha$ profile work we found
very few systems with an obviously low inclination angle (i.e. showing
a strongly split profile) - though this is not something to which the
data were very sensitive.


\section{Conclusions}

The X-ray pulsar population of the SMC is providing an excellent
homogeneous sample of such objects at the same distance and
reddening. Consequently it is possible to investigate the physical
characteristics of such high mass X-ray binaries and put the
conclusions in the context of a differening environment to the
galactic population.
From this work the following conclusions have
been drawn:

\begin{itemize}

\item The spectral class
distribution shows a much greater agreement with those of isolated Be
stars, and appears to be in some disagreement with the galactic
population of Be stars in Be/X-ray binaries. Future accurate 
spectral measurements will provide better determinations of the
spectral class than the photometry presented in this work. 

\item Studies of the long term
optical modulation of the Be star companions reveal an extremely
variable group of objects, a fact which is almost certainly linked 
to the observed large X-ray variability. The common underlying link is the
rate of mass outflow from the Be star and its contribution to the
circumstellar disk.

\item The
spatial distribution of these systems within the SMC strongly suggests
a link between massive star formation and the HI density. Though the
line of sight density needs to be interpretted in three dimensional
terms, the results do point to a clear preference for massive star
formation under specific conditions.

\item Finally, studies of the circumstellar disk characteristics
(equivalent width, optical variability, infra-red colours) 
reveal a strong link with the degree of optical variability offering
potential clues into the long-term stability of such disks.

\end{itemize}

As further observations increase the population of these objects in
the SMC these initial conclusions can be further examained. Given the
current growth rate of discoveries, the final population numbers may
well exceed the galactic population by a factor of ten, and hence they
will becoming increasingly important as reference population for
understanding the evolution of such massive/neutron star binaries.

\section*{Acknowledgments}

This research has made use of the SIMBAD database, operated at CDS,
Strasbourg, France.

This paper utilizes public domain data obtained by the MACHO
Project, jointly funded by the US Department of Energy through the
University of California, Lawrence Livermore National Laboratory
under contract No. W-7405-Eng-48, by the National Science
Foundation through the Center for Particle Astrophysics of the
University of California under cooperative agreement AST-8809616,
and by the Mount Stromlo and Siding Spring Observatory, part of
the Australian National University. 

This research has also made
use of the NASA/ IPAC Infrared Science Archive, which is operated
by the Jet Propulsion Laboratory, California Institute of
Technology, under contract with the National Aeronautics and Space
Administration. 

We are very grateful to Andrzej Udalski for supplying the Optical
Gravitational Lensing Experiment (OGLE) data.

As always, we are remain grateful to the
staff of SAAO for their support during our observations.

JLG acknowledges support for a PhD studentship funded jointly by RAL
and Southampton University. VAM acknowledges support from a SALT
Studentship funded jointly by the NRF of South Africa, the British
Council and Southampton University.

Finally, we are very grateful to the anonymous referee for their
careful and thorough reading of this paper.

\bsp

\label{lastpage}

\end{document}